# Equation Discovery, Parametric Simulation, and Optimization Using the Physics-Informed Neural Network (PINN) Method for the Heat Conduction Problem

Ehsan Ghaderi, Mohamad Ali Bijarchi[1], Siamak Kazemzadeh Hannani, and Ali Nouri-Boroujerdi

Department of Mechanical Engineering, Sharif University of Technology, Tehran, Iran

October 25, 2025

**Abstract**

In this study, the capabilities of the Physics-Informed Neural Network (PINN) method are investigated for three major tasks: modeling, simulation, and optimization in the context of the heat conduction problem. In the modeling phase, the governing equation of heat transfer by conduction is reconstructed through equation discovery using fractional-order derivatives, enabling the identification of the fractional derivative order that best describes the physical behavior. In the simulation phase, the thermal conductivity is treated as a physical parameter, and a parametric simulation is performed to analyze its influence on the temperature field. In the optimization phase, the focus is placed on the inverse problem, where the goal is to infer unknown physical properties from observed data. The effectiveness of the PINN approach is evaluated across these three fundamental engineering problem types and compared against conventional numerical methods. The results demonstrate that although PINNs may not yet outperform traditional numerical solvers in terms of speed and accuracy for forward problems, they offer a powerful and flexible framework for parametric simulation, optimization, and equation discovery, making them highly valuable for inverse and data-driven modeling applications.

**Introduction**

In recent years, the use of artificial intelligence in various branches of science and engineering has expanded rapidly, giving rise to a new research paradigm known as Scientific Machine Learning (SciML) [1-2]. In this emerging field, both physical laws and available data are integrated to solve complex problems more effectively. Traditional numerical methods for thermal and fluid applications, such as finite difference and finite element methods, which rely solely on the governing physical equations, often lack flexibility when applied to diverse or ill-posed problem types [3]. Conversely, purely data-driven artificial intelligence methods, which do not incorporate physical principles, can be impractical in scenarios with insufficient data [4]. Bridging these two approaches, Physics-Informed Neural Networks (PINNs) have been introduced as one of the most powerful frameworks within SciML [5]. In PINNs, the governing equations of the physics, along with any external data or initial/boundary conditions, are embedded directly into the learning process [6]. This integration has led to their successful application in numerous engineering contexts [7-8].

---

[1] corresponding author: bijarchi@sharif.edu

For standard forward problems, where the complete physics is known and the problem is well-posed, the use of PINNs does not necessarily offer significant advantages over conventional numerical methods [9]. However, the strength of PINNs becomes apparent in parametric simulation, where the goal is to obtain solutions for a range of different geometric or physical parameter values [10]. In traditional numerical methods, obtaining solutions for different parameter values typically requires separate, independent executions of the solver. In contrast, the PINN framework can learn this parametric dependence in a parallelized manner within a single training process [10]. Recent studies have demonstrated the successful application of PINNs in parametric simulations across various fields. For instance, Sun et al. [11] investigated the parametric solution of flow around an airfoil, where the angle of attack was incorporated as an input parameter during the network's training, subsequently enabling optimization for the optimal angle of attack. In another study, Cao et al. [12] selected the geometric parameter of an obstacle within a channel for parametric simulation, reporting results for various obstacle dimensions. Ren et al. [13] applied parametric simulation to high Reynolds number flows, incorporating the Mach number as a physical parameter, and demonstrated the method's capability to predict flow behavior across different Mach numbers, with results validated against Computational fluid dynamics (CFD) studies. These examples underscore the growing interest in parametric simulations using PINNs as a flexible and efficient modeling tool. Parametric simulation using PINNs remains an active and attractive area of research [14-17].

Parametric simulation also facilitates optimization, as having model outputs over a range of parameter values naturally enables the identification of optimal conditions. Nevertheless, in many cases, the objective is to infer an unknown physical property or geometric dimension that satisfies a specific condition, framing the problem as an inverse problem [18]. Solving inverse problems, which are often ill-posed, is challenging with conventional CFD methods [3]. The inherent flexibility of PINNs makes them particularly suitable for such tasks, as evidenced by their use in various studies. For example, Arzani et al. [19] highlighted the utility of PINNs in biofluid mechanics applications, where problems are frequently not well-posed, such as cases with unknown boundary conditions (e.g., outflow profiles in blood vessels) or uncertain material properties (e.g., blood viscosity under pathological conditions). In another investigation, Cai et al. [20] used PINNs to identify unknown boundary conditions in a heat transfer problem involving airflow around a coffee cup, combining measured temperature data from thermal imaging with governing physical equations. Similarly, Aliakbari et al. [21] employed concentration data to solve an inverse problem in a channel flow with unknown boundary conditions. These studies collectively highlight the strong potential of PINNs for addressing a wide variety of inverse problems. The application of PINNs to inverse problems across various domains constitutes a vibrant research area, with the cited examples representing only a fraction of the ongoing work [22-23].

Beyond parametric and inverse problems, another area of significant interest is the discovery of governing equations from data obtained via experimental measurements or analogous numerical studies [24-25]. When data is scarce due to experimental limitations or when the underlying physics is partially unknown, a hybrid approach combining data and physical constraints can be highly effective. For instance, Chen et al. [26] utilized limited flow data to infer the coefficients in the general form of fluid dynamic equations. In cases where the precise governing equation is unknown, such as in remaining useful life (RUL) prediction, PINNs have been successfully employed. Liao et al. [27] demonstrated that PINNs could accurately predict RUL by integrating available experimental test data with general knowledge of the governing equation's form. Their results demonstrated that even with incomplete physical knowledge, PINNs can successfully learn the governing relationships. When the exact equation is unspecified, a common strategy involves testing a combination of derivative terms of different integer orders alongside the available data. This approach has been used by Yang et al. [28] for RUL prediction in rolling bearings and by Jang et al. [29] for battery health estimation. A notable observation from these studies is their reliance on integer-order derivatives. However, the use of fractional derivatives, owing to their greater flexibility in modeling complex, unknown physics, is recommended as a promising alternative for such discovery tasks [30-32].

In the present study, we propose a fractional-derivative-based modeling approach within the PINN framework. Instead of using a combination of integer-order derivatives, we employ fractional derivatives and aim to identify their optimal order as part of the modeling process. For the simulation component, the problem is treated parametrically, enabling the trained model to predict solutions for various parameter states after a single solution process. Furthermore, we evaluate the capability of the PINN method in solving an inverse problem by treating a physical property as an unknown. To ensure a cohesive investigation, a heat conduction problem serves as the common test case across all three phases (modeling, simulation, and optimization), with thermal conductivity acting as the representative physical property in both the parametric and inverse problem analyses. A comprehensive review of the literature indicates that a unified study investigating all three problem types using PINNs has not been previously conducted. Moreover, the application of fractional derivatives for equation discovery within the PINN framework represents a novel contribution of this research.

**Section 1. Parametric Simulation**

This section focuses on the parametric simulation of conductive heat transfer, as illustrated in Figure 1, wherein the thermal conductivity of the material is considered as the varying physical parameter across a specified range. In conventional numerical CFD methods, addressing such a problem necessitates implementing a loop structure, sequentially selecting individual thermal conductivity values from the defined range and performing computations for each instance. This procedure is inherently time-consuming. In contrast, within the proposed PINN framework the thermal conductivity parameter is incorporated as an input to the neural network alongside spatial coordinates and is actively involved in the network's learning process. Consequently, by executing the code and the neural network training process once, the PINN method becomes capable of predicting temperature distributions for any value within the parameter range. A notable advantage of the PINN approach is its generalization capability. The model maintains relatively accurate predictive performance even for parameter values outside the training range, which constitutes a further benefit of the introduced methodology. The network training process employed a hybrid optimization strategy, combining the Adam [33] and L-BFGS [34] optimizers. Furthermore, the generation of random collocation points within both the spatial domain and the parameter range was accomplished using the Latin Hypercube Sampling (LHS) [35] technique. In this implementation, the derivatives required for formulating the PINN loss functions were computed via Automatic Differentiation (AD) [36], analogous to the back propagation process in standard neural network training.

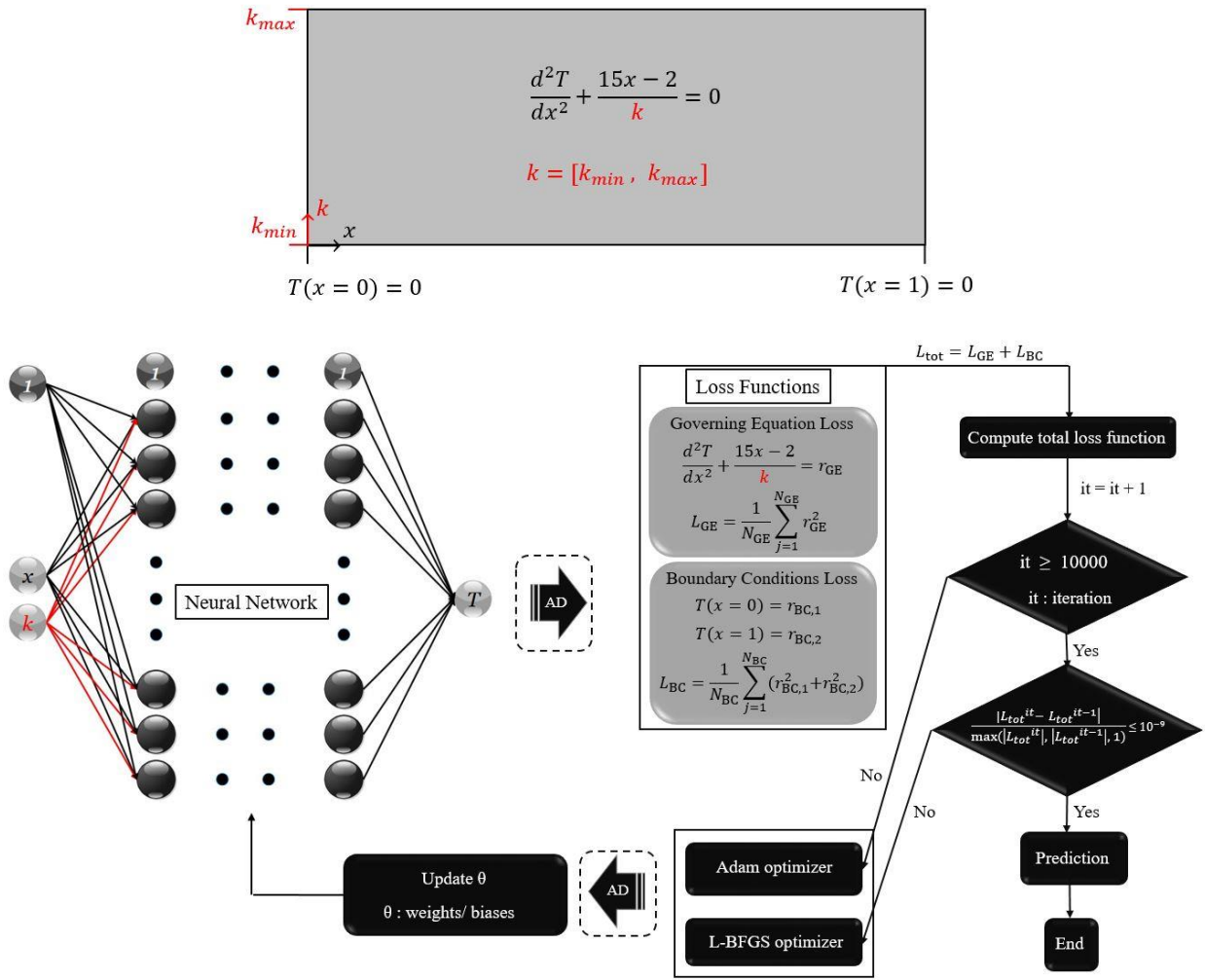

**Figure 1.** Architecture for parametric simulation using the PINN method. (Note: Elements highlighted in red indicate modifications compared to the standard PINN approach).

Figure 2 depicts the temperature distribution along the rod for various thermal conductivity values. It is noteworthy that the model, having been trained only once, successfully predicts the solution for different values of the physical parameter. However, the inclusion of the physical parameter as an additional network input increases the problem's dimensionality, which consequently reduces the computational speed of the proposed method compared to a standard forward PINN problem solving a single physical instance. Another observation is that while the proposed method can yield solutions for parameter values outside the designated training range, the accuracy of these predictions naturally diminishes. The implementation details for this methodology are available in a GitHub repository at the following address: [https://github.com/EhsanGh94/PINN].

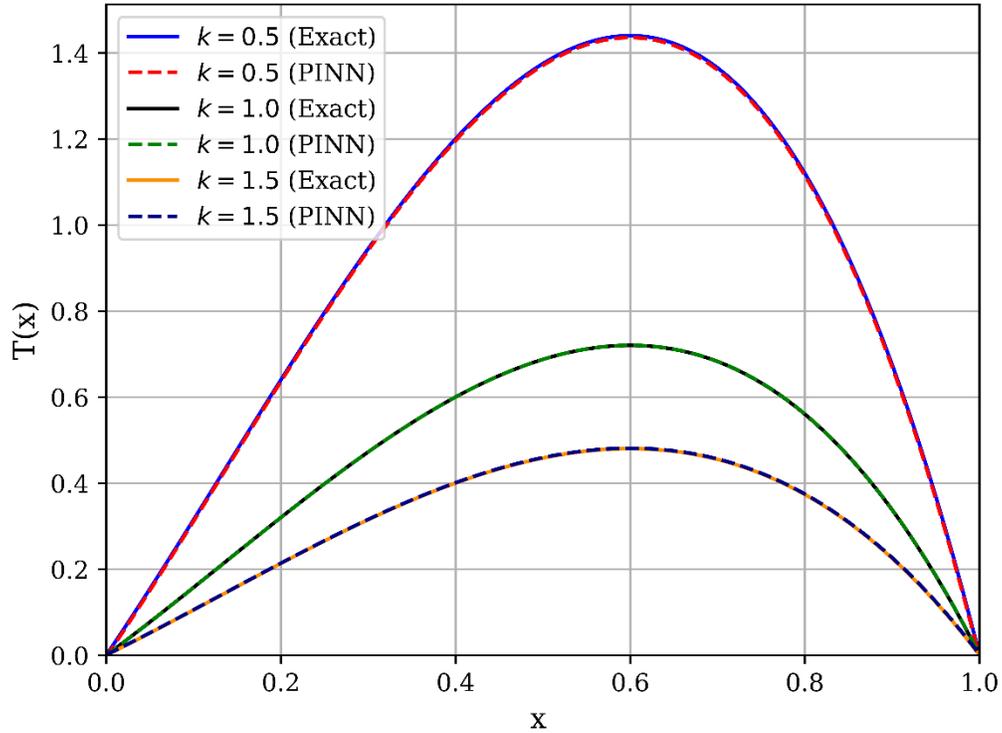

**Figure 2.** Temperature distribution for various thermal conductivity values in the parametric simulation using the PINN method.

## Section 2. Inverse Problem Solving

The focus of this section is on solving an inverse problem to determine an unknown physical property, specifically, the material's thermal conductivity in this study. Figure 3 provides a schematic of the problem, outlining the known and unknown elements of the governing equation. The outline of the unknown thermal conductivity necessitates the availability of temperature distribution data at specific points within the domain for the PINN framework to resolve the problem. As shown in Figure 3, leveraging the availability of a known analytical solution for the governing equation, temperature values from this solution at randomly selected points are incorporated into the network's learning process. This is achieved by adding a data-driven loss term, highlighted in red in the figure. Traditional CFD methods for solving such an inverse problem typically require numerous forward simulations with different assumed thermal conductivity values, a process that is computationally intensive. In contrast, the PINN approach treats the thermal conductivity as an additional unknown parameter, which is concurrently optimized alongside the neural network's weights and biases using the optimization algorithm. Since only a single parameter (the thermal conductivity) is added to the set of network parameters, the overall computational time does not increase significantly.

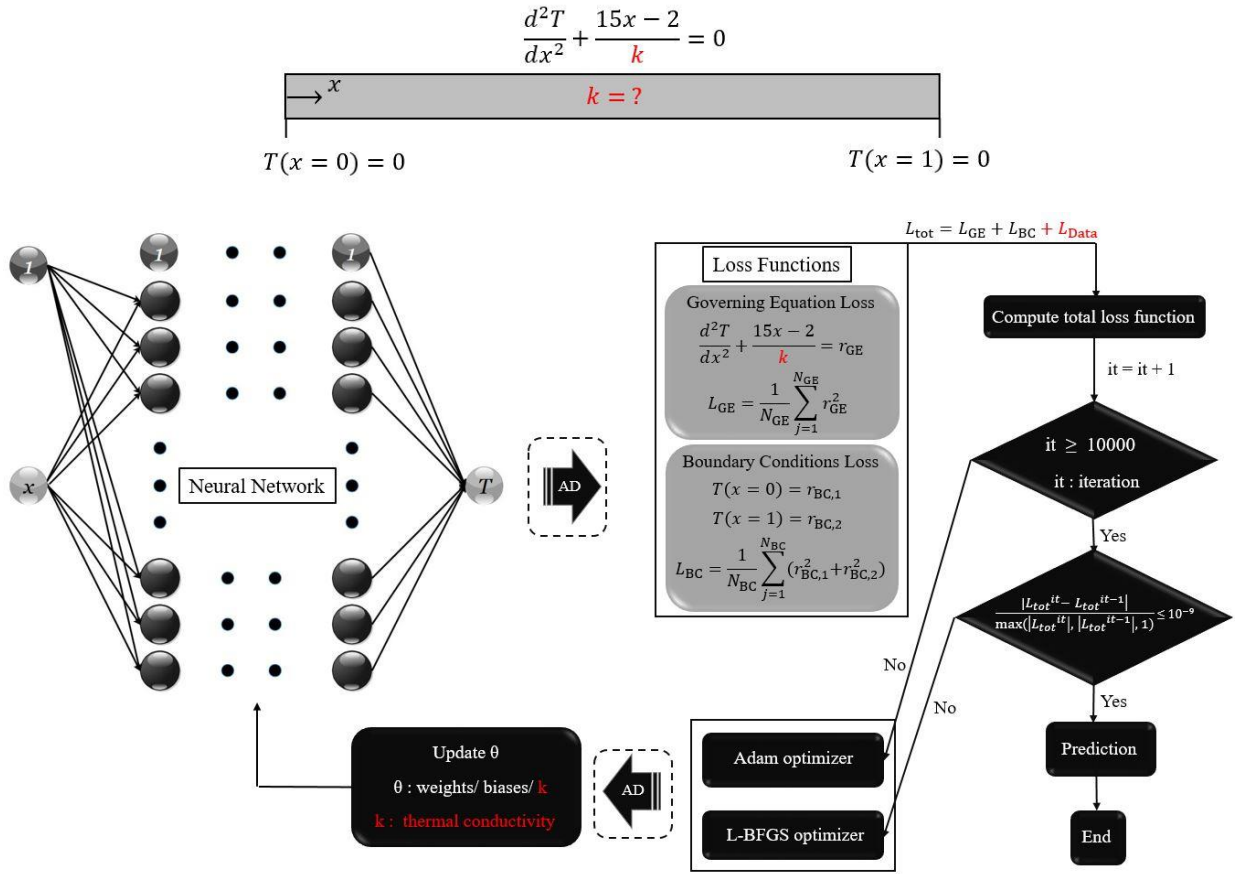

**Figure 3.** Architecture for solving the inverse problem using the PINN method. (Note: Elements highlighted in red indicate modifications relative to the standard PINN approach).

Figure 4 illustrates the estimation of the unknown thermal conductivity using the described PINN framework. The abrupt change in the loss landscape observed around iteration 1000 corresponds to the transition from the first-order Adam optimizer to the second-order L-BFGS optimizer. As evident from the figure, the PINN method successfully converges to the final target value of thermal conductivity within only 1400 iterations. It is important to note that, as with many optimization problems, the initial guess for the parameter significantly influences convergence to the correct solution. If the initial guess deviates substantially (either far too low or high) from the true value, the proposed method may struggle to converge accurately and, in some cases, may even diverge. Nevertheless, the flexibility of the proposed framework is a notable strength. In scenarios where experimental data or numerical results from other sources are available, this method can effectively integrate that data with the underlying physics of the problem, demonstrating its potential for solving complex engineering challenges.

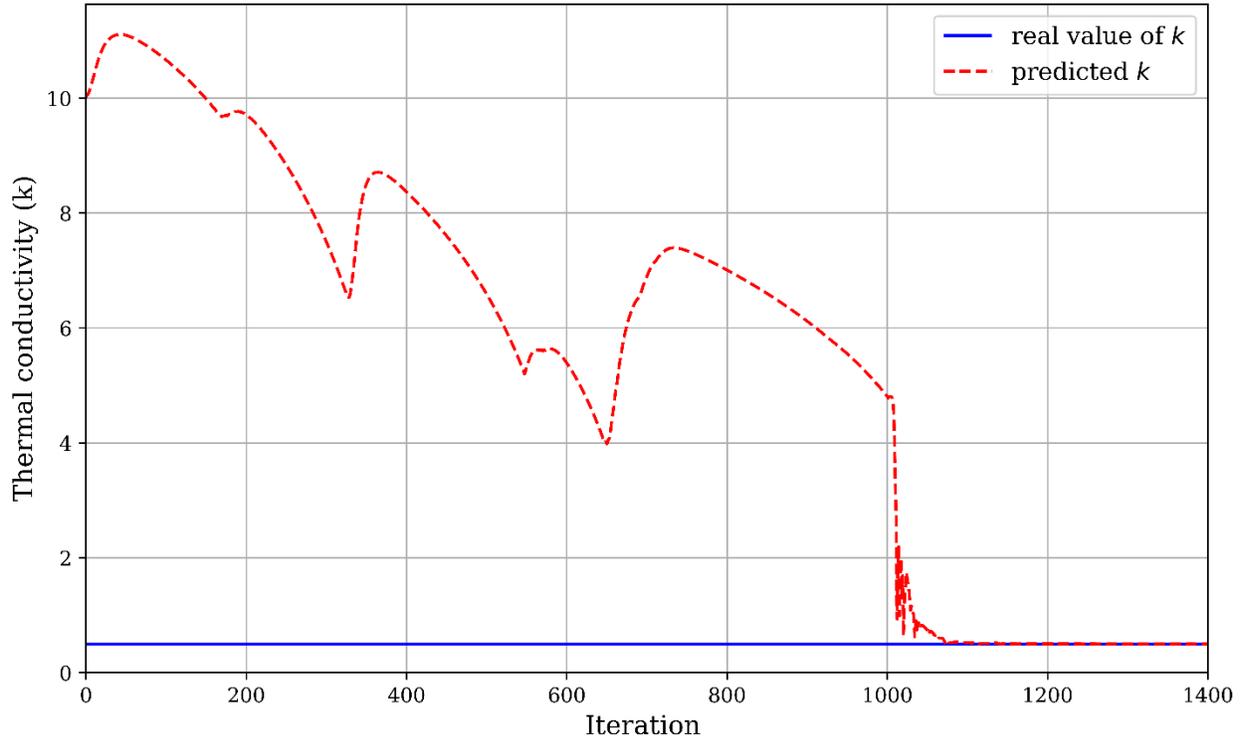

**Figure 4.** Estimation of thermal conductivity in the inverse problem solved via the PINN method.

**Section 3. Governing Equation Discovery**

In the final part of this research, which can be considered an extension of the previous section, we focus on discovering the governing equation of the problem by leveraging a general understanding of the physics and available data. As illustrated in Figure 5, the governing equation is generally assumed to be a fractional differential equation with an unknown derivative order. Recent studies have explored equation discovery using combinations of integer-order derivatives [37-38]. Here, instead of using a combination of integer-order terms, we attempt to discover the governing equation describing the physical behavior using a single derivative term, a fractional derivative. Due to the presence of the fractional derivative, the use of standard Automatic Differentiation (AD) is not feasible in this section. Instead, as shown in Figure 5, fractional derivatives must be computed according to definitions such as Caputo or Riemann-Liouville, employing numerical discretization schemes [39-41]. As can be inferred from the neural network structure in the figure, a data loss term is also incorporated in this framework. For simplicity, these data are injected into the problem by using the exact solution of the equation at random points within the domain. The process of discovering the fractional order, $\alpha$, is treated as an inverse problem, similar to the previous section. Consequently, the value of $\alpha$ is obtained alongside the neural network's weights and biases during the optimization process.

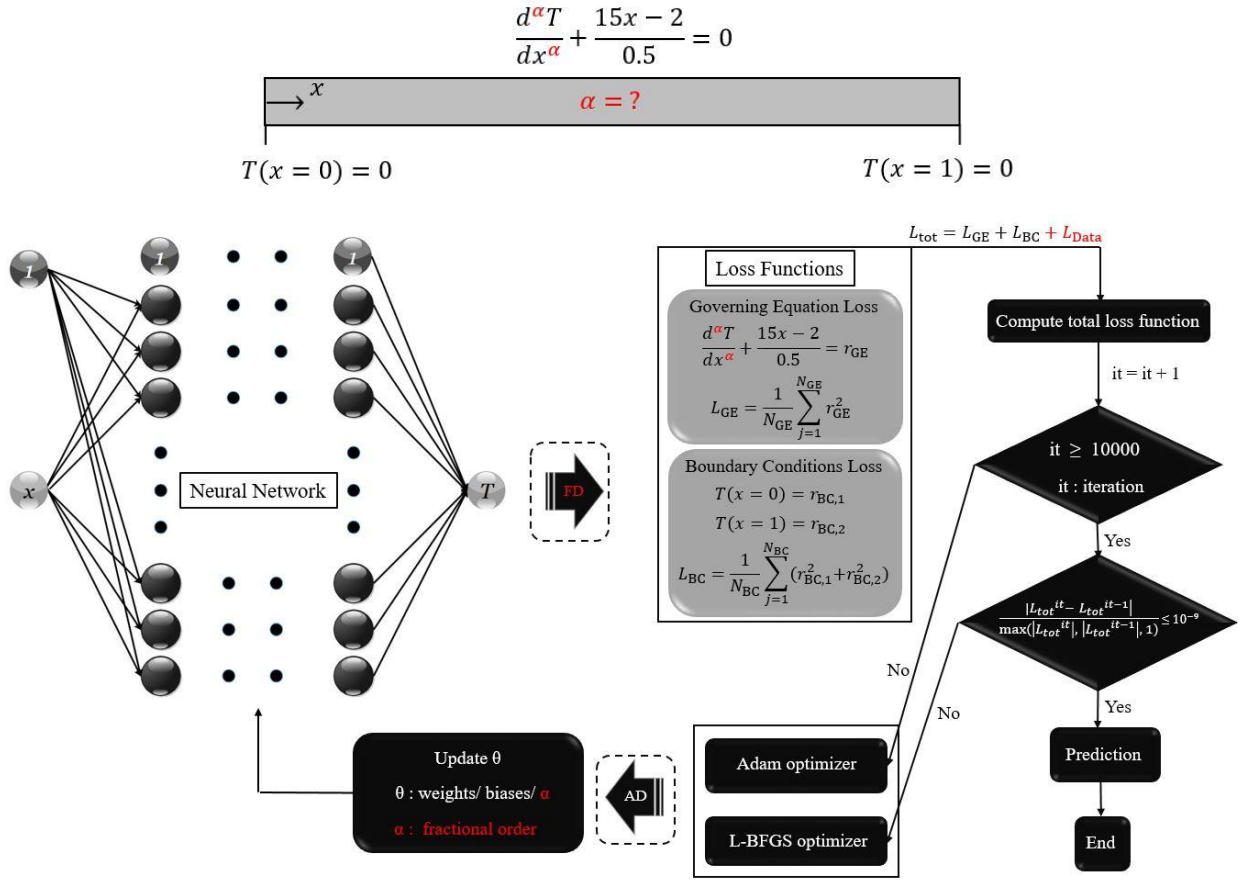

**Figure 5.** Architecture for governing equation discovery using the PINN method. (Note: Elements highlighted in red indicate modifications relative to the standard PINN approach).

Figure 6 shows the solution obtained using the fractional derivative PINN method alongside the true solution of the governing equation, demonstrating the high accuracy of the proposed method. It is important to note that, given the optimization-based nature of the method, finding the correct fractional order is highly dependent on the initial guess. Specifically, initial guesses for $\alpha$ smaller than 0.5 or larger than 3.5 lead to divergence of the method. The introduced framework demonstrates strong generalization potential and can be effectively utilized for the discovery and estimation of governing equations in other applications. The implementation details for this methodology are available in a GitHub repository at the following address: [https://github.com/EhsanGh94/PINN].

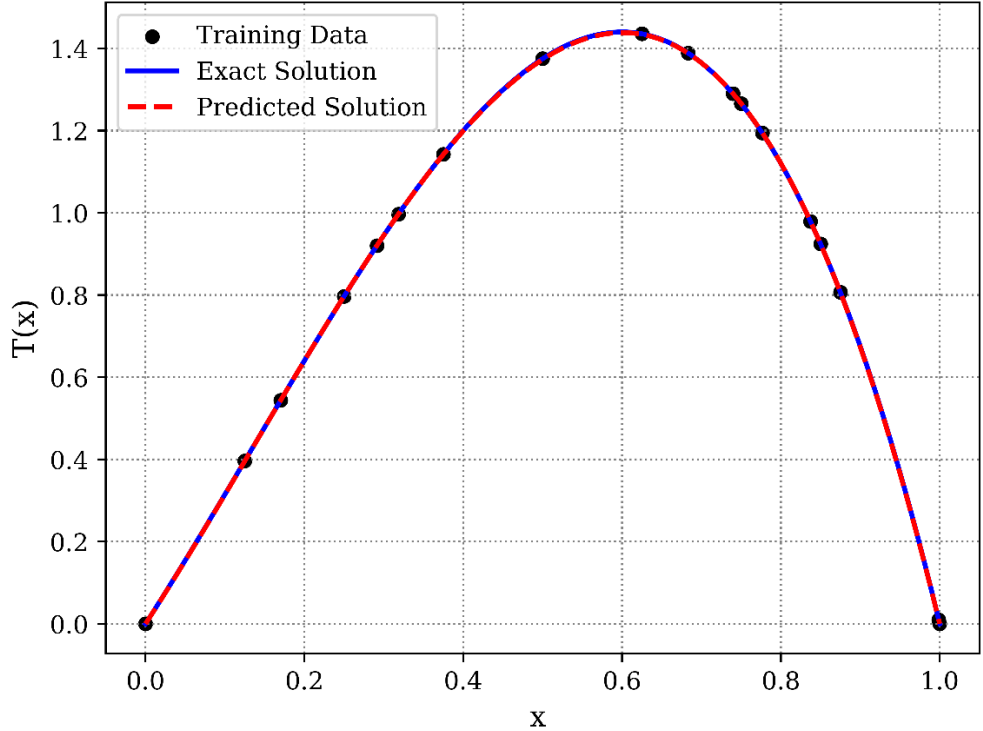

**Figure 6.** Temperature distribution obtained using the PINN method for the fractional differential equation.

**Conclusion**

This study has presented a comprehensive investigation into the application of the Physics-Informed Neural Network (PINN) method across three key areas (modeling, simulation, and optimization) with a specific focus on conductive heat transfer. The principal findings for each section can be summarized as follows:

1. **In the simulation section**, leveraging the parametric simulation capability of PINNs, the method successfully predicted temperature distributions for various thermal conductivity values after a single execution of the training process.

2. **In the optimization (inverse problem) section**, the PINN framework effectively identified the unknown physical property of thermal conductivity by utilizing known temperature data at specific points within the solution domain.

3. **In the modeling section**, where the order of the governing differential equation was considered unknown and potentially fractional, the PINN method demonstrated a robust capability for accurately discovering this parameter and solving the associated problem.

In summary, given recent advancements in artificial intelligence and the ongoing research dedicated to enhancing PINN methodologies, it is anticipated that in the near future, PINNs will become competitive with conventional numerical methods even for standard forward problems.

Nevertheless, the cases examined in this research already underscore the significant flexibility and substantial potential of the PINN approach in addressing a diverse range of engineering challenges.

**Acknowledgment**

The authors gratefully acknowledge the assistance of AI-based language tools in improving the clarity and fluency of the English writing. All ideas, arguments, and scientific content are solely the responsibility of the authors.

**Statements and Declarations**

No funding was received to assist with the preparation of this manuscript.